\documentclass[preprint,nofootinbib,superscriptaddress]{revtex4}
\usepackage{graphicx}
\usepackage{amsmath} 
\usepackage{verbatim} 
\usepackage{mathrsfs}
\usepackage{amsfonts}
\usepackage{dsfont}
\usepackage[latin1]{inputenc}

\newcommand{\tc}{\eta_{\vec{k}}^c}
\newcommand{\beq}{\begin{equation}}
\newcommand{\eeq}{\end{equation}}
\newcommand{\nk}{\vec{k}}

\begin{document}

\author{Alberto Diez-Tejedor}
\email{alberto.diez@fisica.ugto.mx}
\affiliation{Instituto de Ciencias Nucleares, Universidad Nacional Aut\'onoma de M\'exico, M\'exico D.F. 04510, M\'exico}
\affiliation{Departamento de F\'isica, Divisi\'on de Ciencias e Ingenier\'ias, Campus Le\'on, Universidad de Guanajuato, Le\'on 37150, M\'exico }

\author{Gabriel Le\'on}
\email{gabriel.leon@ts.infn.it}
\affiliation{Instituto de Ciencias Nucleares, Universidad Nacional Aut\'onoma de M\'exico, M\'exico D.F. 04510, M\'exico}
\affiliation{Department of Physics, University of Trieste, Strada Costiera 11, 34014 Trieste, Italy}

\author{Daniel Sudarsky}
\email{sudarsky@nucleares.unam.mx}
\affiliation{Instituto de Ciencias Nucleares, Universidad Nacional Aut\'onoma de M\'exico, M\'exico D.F. 04510, M\'exico}
\affiliation{Instituto de Astronom\'ia y F\'isica del Espacio, Universidad de Buenos Aires, Casilla de Correos 67, Sucursal 28, Buenos Aires 1428, Argentina}

\date{ V.1 \today}

\title{The collapse of the wave function in the joint metric-matter quantization for inflation.}

\begin{abstract}

It has been argued that  the standard inflationary scenario
suffers  from a serious deficiency as a model for the origin of the seeds of cosmic structure:
it can not truly account for the transition from an early homogeneous and isotropic
stage to another one lacking such symmetries.
The issue has   often been   thought  as a standard instance of the
``quantum measurement problem",  but  as has  been recently argued
by some of us, that quagmire reaches a critical level in the cosmological context of interest here. This has lead to a proposal
in which the standard paradigm is supplemented by a hypothesis  concerning the self-induced  dynamical  collapse of the  wave
function,  as representing   the physical mechanism    through  which  such  change of  symmetry  is brought forth.
This proposal  was  originally formulated  within the context of
     semiclassical gravity.
Here we investigate  an alternative  realization of such idea
implemented  directly  within the standard  analysis  in terms of a quantum field   jointly
    describing the inflaton and metric perturbations, the so called  Mukhanov-Sasaki variable.
We  show  that   even though the  prescription is
quite different,
the  theoretical predictions
include  some
deviations from the standard ones,  which  are  indeed very similar to those  found in the early studies.
     We  briefly discuss  the  differences  between the two prescriptions, at both,  the conceptual and practical levels.

\end{abstract}

\maketitle

\section{Introduction}
\label{intro}
Inflation represents one of the central cornerstones of modern cosmology. It was  initially proposed  as a solution  to
the classical naturalness problems of the the big bang model, but its impact became even more significant when it came to be regarded as a natural mechanism
to account for the seeds of cosmic structure. However, as discussed in \cite{Perez:2005gh} (see also \cite{Sudarsky:2009za}),
these claims are not fully justified.
The point is that there  is {\it nothing} that could account for the  transmutation of the homogeneous and isotropic vacuum characterizing
the quantum properties of  the early universe,
into   something  that  might be identified  with the  inhomogeneous  and  anisotropic  state  characterizing the universe at, say,
the last scattering surface,  from which the  cosmic  microwave background (CMB) photons were emitted.
At this point we should  warn the reader that  our posture in this regard is not shared  by  all the people   working in the field, and  thus we   invite  him/her to consider the arguments
  on the two sides    of the issue  by him/her-selves. For  a sample  of   works    expressing   views   contrary to ours, please  see the references given in \cite{Others  Inflation,Mukhanov:1990me,Polarski}.
(Note however that different authors in the sample point to $-$slightly$-$ different schemes, indicating
that each author does not find the schemes espoused by other
colleagues to be fully satisfactory).

The problem  was noted   early on  in \cite{Padmanabhan1993} by Padmanabhan,
but the issue was highlighted in \cite{Perez:2005gh}, where the first proposal to address it was put forward.
More recently,  some books in  which  the standard  
picture is presented have mentioned the problem  explicitly (see for instance
\cite{Weinberg2008,Mukhanov2005,Lyth2009} and \cite{Penrose2005}),
while  other  researchers   still claim that there is no outstanding issue \cite{Polarski}.
The  detailed analysis of such postures and the remaining shortcomings  have been discussed  in \cite{Sudarsky:2009za}, and  the  arguments  will not be reproduced  here.
However, let us note that most researchers in the field, including those who acknowledge that  there is  something missing in the standard picture, are convinced that this is in a sense ``just  the standard interpretational problem of quantum mechanics", and  as  such, the issue is one  of ``pure  philosophy"  without any possible impact on the predictions of the theory.
In fact,  the  issue is  sometimes  presented  as that  of the ``quantum to classical transition",  but this, in our view,  hides  the  seriousness of the real problem. We  need some physical process to account for the passage
from a state   with a certain symmetry (homogeneity and  isotropy)  to another one lacking those symmetries,  in a situation  were we can find no  physical
mechanism that might  account for that.
The point is that by
labelling this  issue as just philosophy, what  is meant  by most physicists  is  the  belief that the final results do not depend on the details of whatever one envisions  as  being behind the process
that leads to the emergence of the primordial inhomogeneities out of the quantum  ``fluctuations" (or, more precisely, {\it uncertainties}).
However, as we have
shown in  previous works \cite{Perez:2005gh}, this is not the case,
and particular aspects  of the  process
could have left some imprints in the distribution of matter
and the CMB.

The  idea  that  has been considered in previous  works  as a possibility to deal with that problem involves  adding  to the standard
model the  hypothesis that  the   collapse of the  wave function is  an actual {\it physical} process that occurs  independently  of
external observers. It was initially proposed in \cite{Perez:2005gh}, and developed further in
\cite{DeUnanue:2008fw,Leon:2010fi,Leon:2010wv,Leon:2011ca,DiezTejedor:2011ci,Leon:2011hs,Landau:2011aa} (see also \cite{Sudarsky:2009za,Sudarsky:2007tp,Sudarsky:2006zx}).
As  discussed  in \cite{Sudarsky:2009za}, something that effectively  might be described  in terms of such a collapse of the wave
function could have its origins in  the passage from the  atemporal regime of quantum gravity  to the classical   space-time
description underlining the  general theory of  relativity. That is, in going from one description to the other,  we  might be forced
to characterize   some aspects of the underlying physics in terms  of  sudden  jumps which are not compatible with the unitary Schr\"odinger evolution,
and  which    modify the  state of the universe in an stochastic  way, being therefore capable of transforming  a condition that was
initially homogeneous and isotropic  into another one that is  not (the interested  reader can  consult  the  above   mentioned works as those   issues are not the focus of  this  paper).

The  point we  want to analyze here is to what extend do  the general aspects and details of the results  obtained in previous  works
depend on the specific approach to deal with the quantum  aspects of the problem.
The fact is that in all  previous treatments  we have relied on what is  known  as  semiclassical  gravity. That is, a {\it classical}  description for the space-time metric  (including its  perturbations) coupled  to  a {\it quantum} treatment of the  inflaton field. 
We considered  a quantum field  evolving  in a space-time background, together with the assumption that
the expectation value of the energy momentum tensor
acts as the source of gravity in  Einstein's theory.
The collapse has  been assumed  to  occur  at the level of the  quantum description of the inflaton field, while  the metric  would
simply   respond  to the modification in the  expectation value of the energy momentum tensor leading to  a geometry that is no longer homogeneous and isotropic.
We have found in that case that the details of the model of collapse  and the times  assumed for the collapse  of the various modes  have an  impact on the details of the CMB  power spectrum \cite{DeUnanue:2008fw}.

In this work  we   will  consider a  similar analysis but implemented within a treatment that considers simultaneously the metric and
the scalar field  perturbations (both treated at the quantum level), while the space-time and inflaton homogeneous and isotropic background
will be treated classically. 
That is, we will describe the system of interest in terms of the so  called Mukhanov-Sasaki variable \cite{Mukhanov:1985,Sasaki:1986}, which will be  quantized  in the standard  way,  as it is now  customary on the  literature on the subject.
However, we  will modify  the standard treatment  with the inclusion of what we believe is  the missing  element,   which  at the phenomenological level would  
correspond to  a dynamical collapse  of the wave function, postulated  as reflecting  a yet  undiscovered  aspect  of  Nature,  perhaps  related to quantum gravity as suggested by Penrose \cite{Penrose:1986} and Di\'{o}si \cite{Diosi:1984}.

Our motivation for this paper is twofold. On the one hand we will present the ``self-collapsing" universe within
what today is considered the standard approach to inflation. We believe this could made our ideas more accessible to the community.
On the other hand, we will show that some  important  conclusions regarding  the modification of the power spectrum  are present  in  the two
different  manners  of  incorporating the collapse hypothesis  into the formalism. The exact form of the modifications will differ from one approach  to the other, so, in principle, we could use the cosmological  observations to infer which one of the two pictures provides an  appropriate effective description for the gravitational interaction  at the quantum-classical interplay: semiclassical gravity or that  proposed by Mukhanov and Sasaki. At this point it is important to emphasize  that  the  issue of which is the   variable
that one  should quantize   is  one  with  physical consequences.  Thus  the
Mukhanov-Sasaki approach is more than a particular  formalism, as it
 leads  to  the  quantization of a particular combination of metric  and   scalar field
perturbations. This   contrast  with  the approach  based on  semiclassical  gravity
where only   the scalar filed is quantized,
 and of course this  can  lead to  differences  between the results obtained in this
paper and those obtained in previous  works   by our group.

 The  manuscript is  organized as follows:  In
Section \ref{standard} we briefly review the standard description of cosmological perturbations (both at the quantum and classical level)
in the inflationary scenario. After that, in Section \ref{beyond} we proceed
to make the quantum-mechanical treatment of the field and metric perturbations within the setting of our proposal, and compare our predictions with the observational results.
Finally we discuss our findings in Section \ref{discussion}.


The  conventions  we will be using include a $(-,+,+,+)$
signature for the space-time metric
and natural units with $c=1$.
We  will use the Planck mass $M_p^2\equiv\hbar^2/8\pi G$ and the Planck time $t_p^2\equiv 8\pi G\hbar$, and follow the notation in reference \cite{Mukhanov2005}.
However, we will work  in the ``conformal Newtonian gauge''  from the  beginning (see the expression (\ref{FRW perturbed}) bellow). The reader should recall that   the corresponding equations
coincide (in form) with those  obtained for their ``gauge invariant counterparts"
in \cite{Mukhanov2005} (for a detailed discussion on our motivation for choosing this particular gauge see the  reference \cite{Leon:2010fi}).

\section{The standard approach: a review}\label{standard}

At the classical level the  inflationary universe is described by Einstein's theory $G_{\mu\nu}=8\pi G T_{\mu\nu}$ together with the  equations of motion for the matter fields.
We shall restrict ourselves to the simplest inflationary model,
with a single  scalar field with the standard kinetic and potential terms, the inflaton $\phi$.
We will be only interested in  those configurations very close to a homogeneous and isotropic  Robertson-Walker cosmology.

The study  of  the seeds of cosmic structure depends essentially on the scalar sector of the perturbations.
Ignoring for simplicity
the so called vector and tensor modes,
we can
choose a coordinate system in which
the space-time metric simplifies to
\begin{equation}
ds^{2}=a^{2}(\eta)\left[-(1+2\psi)d\eta^{2}+(1-2\psi)\delta_{ij}dx^{i}dx^{j}\right],\quad\textrm{with}\quad\psi(\eta,\vec{x})\ll 1\label{FRW perturbed}.
\end{equation}
This choice corresponds to the 
``conformal Newtonian gauge'', with
$\psi$ 
an analogue to the
Newtonian potential and $\eta$ the cosmological time in conformal coordinates.
The spatial coordinates $x^i$  are  the  usual  ``co-moving  spatial coordinates".
Note that $\psi=0$ corresponds to  a spatially flat homogeneous and isotropic Robertson-Walker universe.
We shall restrict our attention to a universe that is  very close to a de Sitter solution, and we will not concern ourselves  with the question of  how this concrete
realization
was  obtained from a particular potential term. 
One then focuses on the background universe ($\psi=0$),  which is characterized  in terms of the conformal expansion rate $\mathcal{H}\equiv \dot{a}/a$ (related to the standard Hubble parameter through $\mathcal{H} = aH$)
and the so-called slow-roll parameter $\epsilon\equiv 1-\dot{\mathcal{H}}/\mathcal{H}^2$.
Here the ``dot"  denotes a
derivative with respect to the conformal time.
For practical reasons
we will  often work up to  the lowest non-vanishing order in $\epsilon$,
 taken it as a small positive constant, i.e. $0\le\epsilon\ll 1$. For $\epsilon=0$ we recover a de Sitter universe, $a_{dS}(\eta)=-1/H \eta$, with $-\infty<\eta< 0$ and $H$ constant.

In order to proceed  one
decomposes the scalar field into a homogeneous and isotropic part $\phi_0(\eta)$ plus a small perturbation,
$\phi(x)=\phi_0(\eta)+\delta \phi(\eta,\vec{x})$.
Working  up to the first order in $\psi$ and $\delta\phi$ 
and defining the new fields
\begin{equation}\label{defs}
u\equiv \frac{a\psi}{4\pi G \dot{\phi}_0},\quad v\equiv a\left(\delta\phi+\frac{\dot{\phi}_0}{\mathcal{H}}\psi\right),
\end{equation}
the $00$ and the $0i$
components of the Einstein equations (the Hamiltonian and momentum constraints) can be casted in the form:

\begin{equation}\label{const.pert}
\Delta u=z\left(\frac{v}{z}\right)^{\cdot}\quad\textrm{and}\quad v=\frac{1}{z}\left(zu\right)^{\cdot},\quad\textrm{with}\quad z\equiv\frac{a\dot{\phi}_0}{\mathcal{H}}.
\end{equation}
These two constraints can be combined  with the  dynamical equations 
resulting  in a  single  equation  involving only the field $v$ and the (given) background universe:
\begin{equation}\label{eq.v}
\ddot{v}-\nabla^2 v -\frac{\ddot{z}}{z}v=0.
\end{equation}

Using Friedmann's equation $\mathcal{H}^2 - \dot{\mathcal{H}} = 4\pi G \dot{\phi}_0^2$ and the definition of $\epsilon$
we can re-express $z$ as $z=a\sqrt{\epsilon/4\pi G}$, thus $\ddot{z}/z\simeq\ddot{a}/a$ provided we are in the slow-roll regime. 
At this level the new field $v(\eta,\vec{x})$ contains all the information about the perturbed universe:
the perturbations in the metric and the scalar field can be read from that field
using the constraints (\ref{const.pert}) and the definitions given in (\ref{defs}).
For practical reasons it will be more convenient
to work with periodic boundary conditions over a box of size $L$. We shall take the limit $L\to \infty$ at the end of the calculations.
We can then use a Fourier  decomposition and write
\begin{equation}
v(\eta,\vec{x})=\frac{1}{L^{3/2}}\sum_{\vec{k}\neq 0} v_{\vec{k}}(\eta)\,e^{i\vec{k}\cdot\vec{x}},
\end{equation}
where $k_n=2\pi j_n/L$, $j_n=0,\pm 1,\pm 2,\ldots$, and $n=1,2,3$.
Note that the zero mode has been removed from the perturbed fields.
Recall  that we are working in co-moving coordinates, where the $\vec{k}$'s are fixed in time 
(and related to their physical values through $\vec{k}/a$).
In terms of this decomposition the dynamical equation for  each mode $v_{\vec{k}}(\eta)$ takes the form
\begin{equation}\label{eq.v.modes}
\ddot{v}_{\vec{k}}+\left(k^2-\frac{\ddot{z}}{z}\right)v_{\vec{k}}=0.
\end{equation}

In what follows, and as it is usual in the field, we will consider a classical description for the background universe. However, we shall quantize 
the field $v(x)$ characterizing the small perturbations around the previous symmetric solution. That field can be described in terms of the canonical (first order) action 
\begin{equation}\label{action}
S=\frac{1}{2}\int d\eta\, d^3\vec{x}\left(\dot{v}^2+v\Delta v +\frac{\ddot{z}}{z}v^2\right).
\end{equation}
(See Section 10 in reference \cite{Mukhanov:1990me} for more details.) Recall  that we are working to the linear order
in $\psi$ and $\delta\phi$, so the interaction terms have not been retained here.
At the quantum level the field $v(x)$ and its conjugate momentum $\pi(x)=\dot{v}(x)$ should be promoted to field operators
acting on a Hilbert space $\mathscr{H}$. These operators must satisfy the standard equal time commutation relations
\begin{equation}\label{standard.comm}
[\hat{v}(\eta,\vec{x}),\hat{\pi}(\eta,\vec{y})]=i\hbar\delta(\vec{x}-\vec{y}),\quad [\hat{v}(\eta,\vec{x}),\hat{v}(\eta,\vec{y})]=
[\hat{\pi}(\eta,\vec{x}),\hat{\pi}(\eta,\vec{y})]=0.
\end{equation}
The standard way to proceed now is to decompose
$\hat{v}(x)$ in terms of the time-independent creation and annihilation operators
\begin{equation}\label{decomposition}
 \hat{v}(x)=\frac{1}{L^{3/2}}\sum_{\vec{k}\neq 0} \hat{v}_{\vec{k}}(\eta) e^{i\vec{k}\cdot\vec{x}},
\end{equation}
with $\hat{v}_{\vec{k}}(\eta)\equiv \hat{a}_{\vec{k}} v_k(\eta) + \hat{a}^{\dag}_{-\vec{k}} v^*_{k}(\eta) $ belonging to a set of normal modes satisfying the
 classical equation of motion (\ref{eq.v.modes}) and orthonormal with respect to the symplectic product
\begin{equation}\label{condition}
v_{k}(\eta)\dot{v}^{*}_{k}(\eta)-\dot{v}_{k}(\eta)v^*_{k}(\eta)=i\hbar.
\end{equation}
With
these
definitions
 the commutators (\ref{standard.comm}) translate into
\begin{equation}
 [\hat{a}_{\vec{k}},\hat{a}^{\dagger}_{\vec{k}'}]=\delta_{\vec{k}\vec{k}'},\quad [\hat{a}_{\vec{k}},\hat{a}_{\vec{k}'}]=[\hat{a}^{\dagger}_{\vec{k}},\hat{a}^{\dagger}_{\vec{k}'}]=0,
\end{equation}
%
and the  Fock space
can be constructed in the standard way
starting
with the vacuum state
(i.e.  the state  defined by $\hat{a}_{\vec{k}}\vert 0\rangle=0$ for all $\vec{k}$).

The quantum theory is  specified by
an appropriate choice for the set of functions $v_{\vec{k}}(\eta)$.
However, the equations (\ref{eq.v.modes}) and (\ref{condition}) do not fix that set unequivocally.
Following the standard literature on the subject we will assume the
Bunch-Davies construction, based on functions $v_{k}(\eta)$ that in the asymptotic past
contain only ``positive energy solutions", i.e. for $-k\eta\rightarrow \infty$,
$\dot{v}_{\vec{k}}=-i\omega_k v_{\vec{k}}$ with $\omega_k>0$.
 For a universe close to a de Sitter solution and working to the lowest non-vanishing order in the slow-roll parameter one obtains
\begin{equation}\label{B-D}
v_{k}(\eta)=\sqrt\frac{\hbar}{2k}\left(1-\frac{i}{k\eta}\right)e^{-ik\eta}.
\end{equation}
For that simple construction the state $\vert 0\rangle$ is  known as  the  Bunch-Davies (BD) vacuum.\footnote{Strictly speaking that state is not the BD vacuum, simply because  as the result of a slow-rolling inflaton field the space-time  background cannot be exactly de Sitter. But here we  will ignore that issue
and refer to the state $\vert 0\rangle$ as the BD vacuum, as it is often done.
The relevant issue is that this state is as homogeneous and
isotropic as the true BD vacuum.}
After a few  ``e-folds" of inflation, this
is  expected to  accurately characterize  the state of the inflaton field, and its quantum fluctuations (uncertainties). As we have indicated these are  {\it supposed} to represent the
seeds of
cosmic structure. 
 However, it is  easy to  see that this state is  perfectly homogeneous  and isotropic.\footnote{It is invariant under spatial translations $\hat{T}(d_i)=\exp[i\hat{P}_i d_i]$ and rotations
$\hat{R}_x(\theta_i)=\exp[i\hat{L}(x)_i \theta_i]$, with $\hat{P}_i$ and $\hat{L}(x)_i$ the linear and the angular momentum operators,
and $d_i$ and $\theta_i$ parameters labelling the transformations.}
Thus, a universe  characterized by that state will be homogeneous and isotropic not only at the classical, but also at the quantum level.
Indeed, this is not a surprising result: if the initial state have a symmetry, and the dynamical evolution
preserves that symmetry,\footnote{It  is straightforward to see that the evolution  Hamiltonian commutes with the operators $\hat{T}(d_i)$ and $\hat{R}_x(\theta_i)$.}
the state of the system will be symmetric at any time, and there is nothing (e.g., decoherence, horizon crossing, etc.) that the standard unitary evolution of the quantum
theory could do to avoid  that conclusion. Of course, the standard  accounts  need to  bypass this  \emph{no-go} result  in some
way  or another. However, as it has been argued in detail in \cite{Sudarsky:2009za}, all those attempts   fail to  provide  a  satisfactory answer to
 the question: what is the physical  mechanism  whereby
 the initial symmetry  was  lost?

\section{Beyond the standard quantum theory}\label{beyond}

We  believe that
something
beyond the standard quantum theory, and which we  have previously called ``the collapse",
\begin{equation}\label{collapse}
\vert\textrm{symmetric}\rangle\to \vert\textrm{non-symmetric}\rangle,
\end{equation}
seems to be required in order to break the ``initial" symmetry characterizing the BD vacuum. That process
is  thought to  represent some novel aspect of physics,  connected perhaps  with otherwise  unexpected
properties of quantum gravity, as has  been previously suggested
 by  Di\'osi and Penrose (see also the discussion in Section \ref{discussion}).
In fact, we  should  mention that there  is a long history of
  studies  about  proposals  involving  something like a collapse of  the wave function, emerging basically from the
  community working in  foundations of  quantum theory (see for instance \cite{Bassi:2003gd} and references therein). However,
  those  had never been considered in the present context before the  work
  \cite{Perez:2005gh}.

For the purposes of this paper we will concentrate on the simplest possible case, where
there is only one collapse per mode $\vec{k}$
(of course there is no reason a priori to consider that there could not be more than
one collapse per mode $\vec{k}$, but this is not going to affect our findings, as has been  discussed  in \cite{Leon:2010wv}).
We want to  consider each of the  modes  of the field individually, and  will be assuming that at  $\eta_{\vec{k}}^c$ the mode $\vec{k}$ suffers a collapse, $\vert 0_{\vec{k}}\rangle \to \vert \Xi_{\vec{k}}\rangle$, with $\vert \Xi_{\vec{k}}\rangle$, in principle, some  arbitrary  state chosen from a suitable subset of $\mathscr{H}_{\vec{k}}$.\footnote{In an abuse  of notation we will be writing  
$\mathscr{H}=\prod_{\vec{k}}\mathscr{H}_{\vec{k}}$, although  technically  the   fact that we are   working  with an  infinite  number of degrees of freedom requires  
a construction known as the Fock space. The point  is that, despite not being totally precise, this  way  of presenting things is  more transparent.}
That  is, in principle, we  will allow the collapsing time $\eta_{\vec{k}}^c$ to vary from mode to mode. 
 These collapses will be  assumed to take place according to certain specific  rules which we  will  present in more detail  shortly and which, as we  will see,  will  depend on the particular
 {\it ``collapse  scheme"}
 considered. Each  one of  the collapses will  thus induce a change in the expectation value of the operator $\hat{\psi}_{\vec{k}}(\eta)$.
The point is that after the collapse of the mode  $\vec{k}$,  the universe will be no longer homogeneous and isotropic
(in general) at that  particular  corresponding scale (or more precisely, in regards to that mode).

At this point we  must face the relation between  the quantum
and classical descriptions.
In this particular setting we will focus our attention to
the Newtonian potential, that is,  the scalar metric perturbation  $\psi$ determining the small anisotropies in the temperature of the CMB radiation on the celestial two-sphere
(see the expression (\ref{deltaT}) bellow).
 We  must relate  that perturbation  with the field operator $\hat\psi$ provided  by the
 quantum theory in the previous    discussion.\footnote{
 If we had relied on semiclassical gravity this issue
 did not arise  at this point.
 There the Newtonian potential is a classical
 quantity and the classical to quantum connection occurs at the level of the  Einstein  equations,  where one  side
  is the classical  Einstein tensor  while the
  other side  is the expectation value of the  quantum  energy-momentum operator, i.e. $G_{\mu\nu}=8\pi G \langle \hat{T}_{\mu\nu} \rangle$.}
The connection will be made by taking the view that the classical description is only relevant for those particular states for which the quantity in question is  sharply peaked,
and that the classical description corresponds to the expectation value of said quantity.
We can think for instance in the wave packet of a free particle
  where the  wave function is  sharply peaked around some   position, and that  in  such a case  we could naturally say
  that the  particle's  position is  well defined and  corresponds to  the expectation value of the position operator in  that wave  packet state.
In other words,  we  will be  using the identification
  \begin{equation}\label{psi}
  \psi (x) =
   \langle \Xi |\hat  \psi (x) |\Xi    \rangle,
  \end{equation}
with  $|\Xi    \rangle $ a state of the quantum field $\hat{v}(x)$ characterizing jointly the  metric and field   perturbation,
which of course will be  meaningful  only as long as the  state  corresponds to a sharply peaked  one in the associated variable
$\hat{\psi}(x)$. As we  have noted, if we  consider that the relevant state is  the BD vacuum, as it is usually done,  we  would have a serious problem simply because
$\langle 0 |\hat  \psi (x) | 0 \rangle =0$. This illustrates why  one  is  lead   to  introduce  the collapse  hypothesis.
      In  order to  emphasize the dependence of  the  Newtonian potential on the quantum  state
      we  will often write
       $\psi^{\Xi}(x)$ 
       (in fact, we will generalize this notation to any operator, 
       $\mathcal{O}^\Xi \equiv \langle \Xi | \hat{\mathcal{O}}  |\Xi    \rangle$).
 From the first equation in (\ref{const.pert}) we obtain (in Fourier space):
\begin{equation}\label{source.Psi}
{\psi}^\Xi_{\vec{k}} (\eta) \equiv  \langle \Xi | \hat \psi_{\vec{k}} (\eta)  |\Xi    \rangle =-\frac{\sqrt{4\pi G\epsilon}H}{k^2}\left({\pi}^\Xi_{\vec{k}} (\eta) -\frac{\dot{z}}{z}{v}^\Xi_{\vec{k}} (\eta) \right).
\end{equation}
This is a relation between expectation values of mode operators.
We will be  assuming  that at time $\tc$ a collapse  in the  state of the mode  $\vec k$ has occurred, resulting in  a state characterized (in part)  by  
 the expectation  values of the operators   $ \hat  v_{\vec k} (\eta)$ and   $ \hat \pi_{\vec k} (\eta)$  at that particular time.  We  can then make  use of the
  Ehrenfest's theorem to relate those values  to the expectation values of said operators at any  future time (assuming there  are no additional collapses for that  mode).  In  the present case  those  relations take the form
%
%
%
%
%
%
\begin{subequations}\label{v.y.pi}
\begin{eqnarray}
{\pi}^\Xi_{\vec{k}} (\eta) &=& A_{\vec{k}}(\eta,\eta_{\vec{k}}^c) {\pi}^\Xi_{\vec{k}} (\eta_{\vec{k}}^c) +kB_{\vec{k}}(\eta,\eta_{\vec{k}}^c) {v}^\Xi_{\vec{k}} (\eta_{\vec{k}}^c),\\
 {v}^\Xi_{\vec{k}} (\eta)  &=& k^{-1}C_{\vec{k}}(\eta,\eta_{\vec{k}}^c) {\pi}^\Xi_{\vec{k}} (\eta_{\vec{k}}^c) +D_{\vec{k}}(\eta,\eta_{\vec{k}}^c) {v}^\Xi_{\vec{k}} (\eta_{\vec{k}}^c),
\end{eqnarray}
\end{subequations}
with $A_{\vec{k}}(\eta,\eta_{\vec{k}}^c)$, $B_{\vec{k}}(\eta,\eta_{\vec{k}}^c)$, $C_{\vec{k}}(\eta,\eta_{\vec{k}}^c)$ and $D_{\vec{k}}(\eta,\eta_{\vec{k}}^c)$ some dimensionless functions depending on $k\eta$ and $k\eta_{\vec{k}}^c$. They are not very revealing so we will not write them here explicitly. The interested reader can find these functions in Appendix \ref{A,B,D,D}.
 Using the  expressions  above  and the fact that  for the situation  of interest $\dot z/z  = -(1+\epsilon)/\eta$, we can re-express  $\psi^\Xi_{\vec{k}} (\eta)$ to the lowest non-vanishing order in $\epsilon$ in the   relatively simple form
\begin{eqnarray}\label{Gamma.evolve}
{\psi}^\Xi_{\vec{k}}(\eta)&=&-\frac{\sqrt{4\pi G\epsilon}H}{k^2}\left\{
\pi^{\Xi}_{\vec{k}}(s_{\vec{k}}^c)\left[\cos\Delta_{\vec{k}}^c +\frac{\sin\Delta_{\vec{k}}^c}{s_{\vec{k}}^c}\right]\right.\nonumber\\
&&\hspace{2.5cm}\left.+v^{\Xi}_{\vec{k}}(s_{\vec{k}}^c)k\left[\frac{\cos\Delta_{\vec{k}}^c}{s_{\vec{k}}^c}+\left(\frac{1}{(s_{\vec{k}}^{c})^2}-1\right)\sin\Delta_{\vec{k}}^c\right]\right\}.
\end{eqnarray}
Here we have defined $s\equiv k\eta$, $s_{\vec{k}}^c\equiv k\eta_{\vec{k}}^c$ and $\Delta_{\vec{k}}^c\equiv s-s_{\vec{k}}^c$. Note that, by definition $-\infty <s,s_{\vec{k}}^c < 0$, with $s_{\vec{k}}^c < s$, and then $\Delta_{\vec{k}}^c$ positive definite.

In order to make contact with the observations we shall relate the expression (\ref{Gamma.evolve}) for
the Newtonian potential (only valid during inflation) to
%
%
the small anisotropies observed in the temperature of the CMB radiation, $\delta T(\theta,\varphi)/T_0$. 
They are considered as the fingerprints of the small perturbations pervading the universe at the time of decoupling, and undoubtedly any model for the origin of the seeds of cosmic structure should account for them.
%
%
%
These  data can  be  described  in terms the coefficients  $\alpha_{lm}$
of the multipolar series expansion
\begin{equation}\label{expansion.alpha}
\frac{\delta T}{T_0}(\theta,\varphi)=\sum_{lm}\alpha_{lm}Y_{lm}(\theta,\varphi),\quad
\alpha_{lm}=\int \frac{\delta T}{T_0}(\theta,\varphi)Y^*_{lm}(\theta,\varphi)d\Omega .
\end{equation}
Here $\theta$ and $\varphi$ are the coordinates on the celestial two-sphere, with $Y_{lm}(\theta,\varphi)$ the spherical harmonics ($l=0,1,2\ldots$ and $-l\le m\le l$), and $T_0\simeq 2.725 K$ the temperature average. The different multipole numbers $l$ correspond to different angular scales; low $l$ to large scales and high $l$ to small scales.
%
%
%
%
At large angular scales ($l \leq 20$) the Sachs-Wolfe effect is the predominant source to the anisotropies in the CMB.
That effect relates the anisotropies in the temperature observed today on the celestial two-sphere to the inhomogeneities in the Newtonian potential on the last scattering surface,
\begin{equation}\label{deltaT}
\frac{\delta T}{T_0} (\theta,\varphi) = \frac{1}{3} \psi (\eta_D, \vec{x}_D).
\end{equation}
%
Here  $\vec{x}_D = R_D (\sin \theta \sin \varphi, \sin \theta \cos \varphi, \cos \theta)$, with $R_D$ the radius of the
last scattering surface, $R_D \simeq 4000$ Mpc, and $\eta_D$ is the conformal
time of decoupling. The Newtonian potential
can be expanded in Fourier modes  leading to  $\psi^{} (\eta_D, \vec{x}_D) =
 \sum_{\vec{k} } \psi^{}_{\vec{k}} (\eta_D)\,e^{i \vec{k} \cdot \vec{x}_D}/L^{3/2}$. Furthermore, using
 that $e^{i \vec{k} \cdot \vec{x}_D} = 4 \pi \sum_{lm} i^l j_l (kR_D) Y_{lm} (\theta, \varphi) Y_{lm}^* (\hat{k})$, the expression (\ref{expansion.alpha}) for $\alpha_{lm}$ can be rewritten in the form
\begin{equation}\label{alm2}
\alpha_{lm} = \frac{4 \pi i^l }{3L^{3/2}} \sum_{\vec{k}} j_l (kR_D) Y_{lm}^* (\hat{k})T_{{k}}(\eta_R,\eta_D) \psi^{\Xi}_{\vec{k}} (\eta_R),
\end{equation}
with $j_l (kR_D)$ the spherical Bessel function of order $l$. Here we have included the transfer function $T_{{k}}(\eta_R,\eta_D)$ in order to evolve the perturbation in the Newtonian potential from the end of inflation to the last scattering surface, $\psi_{\vec{k}}(\eta_D)=T_{k}(\eta_R,\eta_D) \psi^{\Xi}_{\vec{k}} (\eta_R)$, with $\eta_R$ the reheating time.
 We  will  be  ignoring this  aspect   from this point onward,  despite the  fact that this  transfer function is  behind  the famous acoustic peaks, the most noteworthy  feature of the CMB power spectrum. The point is that they  relate to aspects of plasma physics that are well understood and thus  uninteresting for our purposes here. This  will mean that 
the observational power spectrum  would  have  such  features removed  before comparing with our results (this is   in the same  spirit  that one removes the 
imprint of our galaxy, 
or the  dipole associated with our peculiar motion). This would be  equivalent  to assume that the observations fits well with a nearly flat Harrison-Zel'dovich  spectrum. 

Note that the expression (\ref{alm2}) has  no analogue in the  usual  treatments of the subject, providing us  with a  clear  identification   of the  aspects  of the  analysis  where  the ``randomness"  is  located.  In this case  it resides  in the  randomly   selected values for 
$\psi_{\vec{k}} (\eta_D)$, i.e. in the randomly selected values for $\psi^{\Xi}_{\vec{k}}$ at the collapsing time, see \eqref{v.y.pi} above. Here we also find a clarification of  how, in spite  of  the intrinsic  randomness,  we  can make  any prediction at all. The  individual  complex quantities  $\alpha_{lm} $  correspond to  large sums of complex  contributions,  each  one  having a  certain randomness,   but leading  in  combination to a characteristic  value  in  just  the same  way as a random walk made  of multiple steps.  Nothing  like  this can be found in the most popular accounts, in which the issues  we have been focusing on here are   hidden  in a maze  of often  unspecified  assumptions and unjustified  identifications \cite{Sudarsky:2009za}.
%
%
More precisely, all the modes $\psi_{\vec{k}} (\eta_D)$ contribute to $\alpha_{lm}$
with a  complex number, leading to what is in effect a sort of ``two-dimensional random walk" whose total displacement corresponds to the  interesting  aspect of the observational quantity (this will be more evident next when we specify the collapse scheme).  It is  therefore clear that, as in the case of any random walk, such quantity can not be evaluated, and the only thing that can be done is to calculate the most likely (ML) value for such total displacement, with the expectation that the observed quantity will be close to that value.  That is,  we  need to  estimate the  most likely value  of 
\begin{equation}\label{alm}
|\alpha_{lm} |^2= \frac{16 \pi^2}{9L^{3}} \sum_{\vec{k}, \vec{k'}} j_l (kR_D) j_l (k'R_D) Y_{lm}^* (\hat{k}) Y_{lm} (\hat{k'})\psi^{\Xi}_{\vec{k}} (\eta_R){\psi^{\Xi}_{\vec{k'}}}^* (\eta_R).
\end{equation}
As it is now standard in our treatments, we do this with the help of an imaginary ensemble of universes (each one corresponding to a possible realization of the collapse), and the identification of the most likely value $|\alpha_{lm}|_{\textrm{ML}}^2$ with the ensemble's mean value.
The  spread of the corresponding  values   within  such   ensemble corresponds to what is usually known as  the cosmic  variance.\footnote{As  such  this quantity can not be  measured,  and is normally  just estimated  for the related quantity
$C_l \equiv \frac{1}{2l+1} \sum_{m } |\alpha_{lm} |^2$ to be given by  the corresponding    Gaussian  value   of  $C_l/\sqrt{l+ 1/2}$.}
It is precisely at this point where there appears the link between the statistics of the quantum theory (we will be assuming that the collapses are guided by the quantum uncertainties) and the statistics over an ensamble of {\it classical inhomogeneous} universes. (In the standard approach all the evolution is unitary and then deterministic).
Under this assumption
we obtain that all the information regarding the ``self-collapsing" universe will be codified in the quantity\footnote{In the standard approach
the $n$-point correlation function for the field operator $\hat{\psi}(x)$ is {\it identified} (without any apparent reason, see for instance the references \cite{Perez:2005gh,Sudarsky:2009za}) with the average over an ensemble of classical anisotropic universes of the same correlation function, now for the Newtonian potential $\psi (x)$. That is the reason for which they identify the expression (\ref{cantidad.importante.gausiana}) with
\begin{equation}\label{cuantico}
 \lim_{-k\eta_{R}\to 0}\langle 0\vert\hat{\psi}_{\vec{k}}(\eta_R)\hat{\psi}^{\dagger}_{\vec{k}'}(\eta_R)\vert 0\rangle=\epsilon\frac{t_p^2H^2}{4k^3}\delta_{\vec{k}\vec{k}'}.
\end{equation}
Note that in the corresponding expressions we have an ensemble average of the product of two 1-point functions in contrast with the 2-point function found in the usual approach.}

\begin{equation}\label{cantidad.importante.gausiana}
 \overline{\psi^{\Xi}_{\vec{k}}(\eta_R)\psi^{\Xi *}_{\vec{k}'}(\eta_R)},
\end{equation}
with the over-bar making reference to the ensemble average: the relevant quantities for the analysis of the seeds of cosmic structure are those  characterizing  the statistics of the collapse. We  will  further  identify this quantity with  the value of the corresponding  limit $-k\eta_R \to 0$, which  can be expected to be  appropriate when  restricting interest  on  the modes that are ``outside the horizon" at the end of inflation, since these are the modes that give a major contribution to the
quantities of observational interest.
 Let us note that the function $\psi^{\Xi}_{\vec{k}}(\eta)$ in (\ref{Gamma.evolve}) depends on the time of collapse, so it is expected that the expression (\ref{cantidad.importante.gausiana}) will also depend (in general) on  the  values of $s_{\vec{k}}^c$.
 As we shall see that will affect the  theoretical  values of the  quantities  of interest, and  in particular the form of  the   power spectrum (see for instance the expression (\ref{power}) bellow).
%
 A  simple and direct  connection  with  the form of the  standard results would be  obtained if  we had
\begin{equation}\label{cantidad.importante.gausiana.esperada}
 \lim_{-k\eta_R\to 0}\overline{\psi^{\Xi}_{\vec{k}}(\eta_R)\psi^{\Xi *}_{\vec{k}'}(\eta_R)}=\frac{2\pi^2}{k^3}\mathcal{P}_{\psi}(k)\delta_{\vec{k}\vec{k}'}, \quad\textrm{with}
 \quad \mathcal{P}_{\psi}(k)=\epsilon\frac{t_p^2H^2}{8\pi^2},
\end{equation}
(see equation (\ref{cuantico}) in footnote).
Here $t_p$ is the Planck time, $H$ the scale of inflation, and $\mathcal{P}_{\psi}(k)$ the power spectrum for the Newtonian potential.\footnote{The standard amplitude for the power spectrum is usually presented as proportional to $V/(\epsilon M_P^4) \propto H^2 t_p^2/\epsilon$, where $V$ is the inflaton's potential. The fact that $\epsilon$ is in the denominator leads, in the standard picture, to a constraint scale for $V$. However, in \eqref{cantidad.importante.gausiana.esperada} the slow-roll parameter $\epsilon$ is in the numerator. This is because
we have not used (and in fact we will not) explicitly the transfer function $T_{k} (\eta_R,\eta_D)$. In the standard literature it is common to find the power spectrum for the quantity $\zeta(x)$, a field representing the curvature perturbation in the co-moving gauge. This quantity is constant for modes ``outside the horizon" (irrespectively of the cosmological epoch), thus it avoids the use of the transfer function. The quantity $\zeta$ can be defined in terms of the Newtonian potential as $\zeta
 \equiv \psi + (2/3)(\mathcal{H}^{-1} \dot{\psi} + \psi)/(1+\omega)$, with $\omega \equiv p/\rho$. For large-scale modes $\zeta_k \simeq \psi_k [ (2/3) (1+\omega)^{-1} + 1]$, and during inflation $1+\omega = (2/3)\epsilon$. For these modes $\zeta_k \simeq \psi_k/\epsilon$ and the power spectrum is $\mathcal{P}_{\zeta}(k) = \mathcal{P}_{\psi}(k) / \epsilon^2 \propto H^2 t_p^2/\epsilon  \propto V/(\epsilon M_P^4)$, which contains the usual amplitude. For a detailed discussion regarding the amplitude within the collapse framework see reference \cite{Leon:2010fi}.}
However, in general, we will not obtain that simple relation. 
This is because the dependence of
(\ref{cantidad.importante.gausiana}) on $s_{\vec{k}}^c$ will break the scale independence of the primordial perturbations, i.e. $\mathcal{P}_{\psi}(k)\neq\textrm{const.}$

\begin{figure}
\begin{center}
  \includegraphics{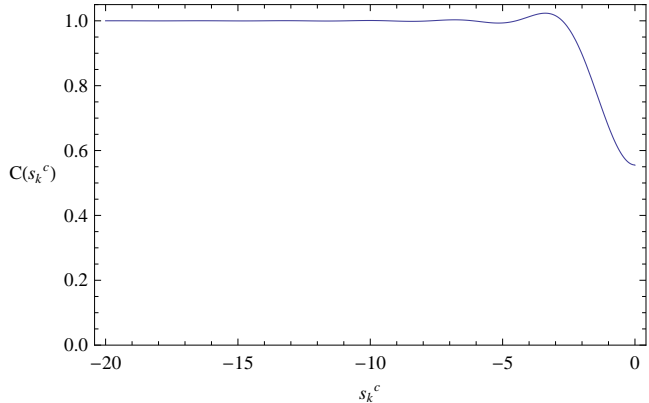}
\end{center}
\caption{The function $C(s_{\nk}^c)$ defined in \eqref{Ck} for $s_{\nk} \in [-20,0)$ using the symmetric scheme $\lambda_v=\lambda_\pi=1$. }
\label{fig:1}       
\end{figure}

In order to see that, let us illustrate our findings with a very simple model for
the collapse. As we  want our collapse process  to  be closely related,  or more  precisely, to mimic the ordinary measurements in standard quantum mechanics, we describe the former in terms of hermitian operators. Thus, we decompose the operators $\hat{v}_{\vec{k}}(\eta)$ and $\hat{\pi}_{\vec{k}}(\eta)$ in their real and imaginary parts, $\hat{v}_{\vec{k}}(\eta)=\hat{v}^{\textrm R}_{\vec{k}}(\eta)+i\hat{v}^{\textrm I}_{\vec{k}}(\eta)$ and $\hat{\pi}_{\vec{k}}(\eta)=\hat{\pi}^{\textrm R}_{\vec{k}}(\eta)+i\hat{\pi}^{\textrm I}_{\vec{k}}(\eta)$,
with $\hat{v}^{\textrm{R,I}}_{\vec{k}}(\eta)=(v_{\vec{k}}(\eta)\hat{a}^{\textrm{R,I}}_{\vec{k}}+v^*_{\vec{k}}(\eta)\hat{a}^{\textrm{ R,I}\dagger}_{\vec{k}})/\sqrt{2}$, $\hat{\pi}^{\textrm{R,I}}_{\vec{k}}(\eta)=(\pi_{\vec{k}}(\eta)\hat{a}^{\textrm{ R,I}}_{\vec{k}}+\pi^*_{\vec{k}}(\eta)\hat{a}^{\textrm{R,I}\dagger}_{\vec{k}})/\sqrt{2}$,
and
\begin{equation}
\hat{a}^{\textrm R}_{\vec{k}}=\frac{1}{\sqrt{2}}\left(\hat{a}_{\vec{k}}+\hat{a}_{-\vec{k}}\right), \quad
\hat{a}^{\textrm I}_{\vec{k}}=\frac{-i}{\sqrt{2}}\left(\hat{a}_{\vec{k}}-\hat{a}_{-\vec{k}}\right).
\end{equation}
With these definitions $\hat{v}^{\textrm{R,I}}_{\vec{k}}(\eta)$ and $\hat{\pi}^{\textrm{R,I}}_{\vec{k}}(\eta)$ are
Hermitian operators (i.e. $\hat{v}^{\textrm{R,I}}_{\vec{k}}(\eta)=\hat{v}^{\textrm{R,I}\dagger}_{\vec{k}}(\eta)$ and $\hat{\pi}^{\textrm{R,I}}_{\vec{k}}(\eta)=\hat{\pi}^{\textrm{R,I}\dagger}_{\vec{k}}(\eta)$), but the commutation relations between $\hat{a}^{\textrm R}_{\vec{k}}$ and $\hat{a}^{\textrm I}_{\vec{k}}$ are
non-standard,
\begin{equation}\label{nonstandard.commutators}
[\hat{a}^{\textrm R}_{\vec{k}},\hat{a}^{\textrm R\dagger}_{\vec{k}'}]= (\delta_{\vec{k},\vec{k}'}+\delta_{\vec{k},-\vec{k}'}),\quad
[\hat{a}^{\textrm I}_{\vec{k}},\hat{a}^{\textrm I\dagger}_{\vec{k}'}]= (\delta_{\vec{k},\vec{k}'}-\delta_{\vec{k},-\vec{k}'}),
\end{equation}
with all the other commutators vanishing. Note that, according to (\ref{nonstandard.commutators}), the modes $\vec{k}$ and $-\vec{k}$ in the previous decomposition are not independent. This will have important consequences later. Now, since $\hat{v}^{\textrm{R,I}}_{\vec{k}}(\eta)$ and $\hat{\pi}^{\textrm{R,I}}_{\vec{k}}(\eta)$ are Hermitian operators, they are susceptible of ``being measured": we will assume, in analogy with standard quantum mechanics, that the collapse is somehow analogous to an imprecise measurement of the operators $\hat{v}^{\textrm{R,I}}_{\vec{k}}(\eta)$ and $\hat{\pi}^{\textrm{R,I}}_{\vec{k}}(\eta)$, and that the final result of the collapse will be guided by the quantum uncertainties,
\begin{subequations}\label{the.model}
\begin{eqnarray}
\langle\hat{\pi}_{\vec{k}}^{\textrm{R,I}}(\eta_{\nk}^c)\rangle_\Xi &=& \lambda_{\pi} x_{\vec{k},\pi}^{\textrm{R,I}}\sqrt{\langle 0\vert\left[\Delta\hat{\pi}_{\vec{k}}(\eta_{\nk}^c)\right]^2\vert 0\rangle}=\frac{\lambda_{\pi} x_{\vec{k},\pi}^{\textrm{R,I}}}{\sqrt{2}}\left|\pi_k(\eta_{\nk}^c)\right| ,  \\
\langle\hat{v}_{\vec{k}}^{\textrm{R,I}}(\eta_{\nk}^c)\rangle_\Xi   &=& \lambda_{v} x_{\vec{k},v}^{\textrm{R,I}}
\sqrt{\langle 0\vert\left[\Delta\hat{v}_{\vec{k}}(\eta_{\nk}^c)\right]^2\vert 0\rangle}= \frac{\lambda_{v} x_{\vec{k},v}^{\textrm{R,I}}}{\sqrt{2}}\left|v_k(\eta_{\nk}^c)\right| ,
\end{eqnarray}
\end{subequations}
with $x_{\vec{k},\pi}^{\textrm{R,I}}$ and $x_{\vec{k},v}^{\textrm{R,I}}$ taken to be  a collection of independent  random numbers selected from a Gaussian distribution centered at zero with unit-spread, and $\lambda_{\pi}$ and $\lambda_{v}$
two real numbers (usually  $0$ or $1$) that allow us to  specify the collapse proposal we  want to consider. The mode $\vec{k}$ of any possible realization of the universe will be described in terms of the  specific numerical  values of $x_{\vec{k},\pi}^{\textrm{R,I}}$ and $x_{\vec{k},v}^{\textrm{R,I}}$.
On the other hand, we will be using  the values for $\lambda_{\pi}$ and $\lambda_{v}$ to characterize  the   different  collapse schemes: i) $\lambda_v=0$, $\lambda_\pi=1$ (corresponding to what  was  called  the  ``Newtonian  scheme"  in the  setting of semiclassical gravity), ii) $\lambda_v=\lambda_\pi=1$ (which  was called the  ``symmetric  scheme"),  or even iii) $\lambda_v=1$, $\lambda_\pi=0$  (a scheme  suggested to  us  by  Prof. R. M. Wald).
%
Introducing the expressions for $\langle\hat{\pi}_{\vec{k}}^{\textrm{R,I}}(\eta_{\nk}^c)\rangle_\Xi$ and $\langle\hat{v}_{\vec{k}}^{\textrm{R,I}}(\eta_{\nk}^c)\rangle_\Xi$ given in (\ref{the.model}) into
(\ref{Gamma.evolve}) and (\ref{cantidad.importante.gausiana}) and taking the limit when $-k\eta_R$ goes to zero we obtain
%
%
%
\beq
\frac{\epsilon t_p^2H^2 }{4k^{3/2}k'^{3/2}}\left[ M_{\nk} M_{\nk'} (  \overline{x_{\nk,v}^{\textrm{R}} x_{\nk',v}^{\textrm{R}}} +  \overline{x_{\nk,v}^{\textrm{I}} x_{\nk',v}^{\textrm{I}}} ) + N_{\nk} N_{\nk'} (  \overline{x_{\nk,\pi}^{\textrm{R}} x_{\nk',\pi}^{\textrm{R}}} +  \overline{x_{\nk,\pi}^{\textrm{I}} x_{\nk',\pi}^{\textrm{I}}} ) \right],
\eeq
with
\begin{subequations}
\begin{eqnarray}
M_{\nk} &\equiv & \lambda_v  \bigg[1 + \frac{1}{(s_{\nk}^c)^2}\bigg]^{\frac{1}{2}}  \bigg[\frac{ \cos s_{\nk}^c}{s_{\nk}^c} - \bigg( \frac{1}{(s_{\nk}^c)^2} - 1 \bigg) \sin s_{\nk}^c \bigg],\\
N_{\nk} &\equiv & \lambda_\pi \bigg[ 1 - \frac{1}{(s_{\nk}^c)^2} + \frac{1}{(s_{\nk}^c)^4} \bigg]^{\frac{1}{2}} \bigg[\cos s_{\nk}^c - \frac{\sin s_{\nk}^c}{s_{\nk}^c}  \bigg].
\end{eqnarray}
\end{subequations}
Here we have made use of the independence among the four sets of random variables $x_{\nk,v}^{\textrm{R}}$, $x_{\nk,v}^{\textrm{I}}$, $x_{\nk,\pi}^{\textrm{R}}$ and $x_{\nk,\pi}^{\textrm{I}}$. However, we need to recall that, within each set, the variables corresponding to $\nk$ and $-\nk$ are not independent. This will be reflected by setting
$\overline{x_{\vec{k},i}^{\textrm{R}}x_{\vec{k}',i}^{\textrm{R}}}=\delta_{\vec{k},\vec{k}'}+\delta_{\vec{k},-\vec{k}'}$ and $\overline{x_{\vec{k},i}^{\textrm{I}}x_{\vec{k}',i}^{\textrm{I}}}=\delta_{\vec{k},\vec{k}'}-\delta_{\vec{k},-\vec{k}'}$ (here $i=\pi,v$), in accordance with the commutators (\ref{nonstandard.commutators}). Writing all these expressions together we conclude
\beq\label{limpsieta}
 \lim_{-k\eta_R\to 0}\overline{\psi^{\Xi}_{\vec{k}}(\eta_R)\psi^{\Xi *}_{\vec{k}'}(\eta_R)} = \epsilon\frac{t_p^2 H^2}{4k^3} (M_{\nk}^2 + N_{\nk}^2) \delta_{\nk\nk'}.
\eeq
Comparing (\ref{limpsieta}) with (\ref{cantidad.importante.gausiana.esperada}) we obtain a power spectrum of the form

\begin{subequations}\label{power}
\begin{equation}\label{ps}
\mathcal{P}_{\psi}(k) \equiv \epsilon\frac{t_p^2H^2}{8\pi^2} C(s_{\vec{k}}^c),
\end{equation}
with the definition

\begin{eqnarray}\label{Ck}
C(s_{\vec{k}}^c) &\equiv&  \lambda_{\pi}^2
\left(1-\frac{1}{(s_{\vec{k}}^c)^2}+\frac{1}{(s_{\vec{k}}^c)^4}\right)\left[\cos s_{\vec{k}}^c-\frac{\sin s_{\vec{k}}^c}{s_{\vec{k}}^c}\right]^2 \nonumber\\
&+&  \lambda_{v}^2 \left(1+\frac{1}{(s_{\vec{k}}^c)^2}\right)\left[\frac{\cos s_{\vec{k}}^c}{s_{\vec{k}}^c}-\left(\frac{1}{(s_{\vec{k}}^{c})^2}-1\right)\sin s_{\vec{k}}^c\right]^2.
\end{eqnarray}
\end{subequations}
Note that, in general, the power spectrum
is {\it not} flat, i.e. it depends on $\vec{k}$ through the previously defined quantity $s_{\vec{k}}^c \equiv k \eta_{\nk}^c$. That is, the dependance on $\nk$ in the spectrum is given by the function $C(s_{\nk}^c)$ (see for instance Fig. \ref{fig:1} for the particular case with $\lambda_v=\lambda_\pi=1$).
We will obtain a flat power spectrum if $s_{\vec{k}}^c=\textrm{const}$, that is, if the times of collapse satisfy $\tc = A k^{-1}$, with $A$ a (dimensionless) negative definite constant.
This  is  a  non-trivial  relationship   which could be taken as  providing clues  about the  Nature of the  mechanism behind the collapse.
In reference \cite{Perez:2005gh}, for instance,   it was  shown that a simple   generalization of  a   proposal   by  R. Penrose  (involving aspects
of  what he believes  should be  some   features of quantum gravity)
 would lead  exactly to this simple rule  for the times  of collapse. 
Without  that, we  do not  obtain  the  standard  Harrison-Zel'dovich shape  of the power spectrum   in any of the  simple  recipes  for the  collapse scheme considered, i) $\lambda_v=0$, $\lambda_\pi=1$, ii) $\lambda_v=\lambda_\pi=1$,  or even iii) $\lambda_v=1$, $\lambda_\pi=0$.
In fact, it is  easy to see that  in the absence of  such  specific  pattern for the collapsing time,
it is impossible to adjust $\lambda_v$ and $\lambda_\pi$ in order to recover an exactly flat power spectrum, i.e. to  adjust  those  values  in a way that  the expression $\eqref{Ck}$  becomes  independent of $s_{\vec{k}}^c$. In other words, a  constant function (i.e. independent of $s_{\vec{k}}^c$) and the functions appearing as coefficients accompanying $\lambda_v^2$ and $\lambda_\pi^2$ form a set  of  linearly independent functions. However, we  will see  that  in certain cases  a {\it nearly} flat power spectrum is  possible, even with general patterns for the collapsing time,  as long as
they occur well outside  the Hubble radius, see expressions \eqref{ps-infty} and \eqref{pslim} bellow.

On the other hand, we do not presume that a collapse theory involving stochastic components will lead to
decay times that precisely follow the pattern $\eta_{\nk}^c = Ak^{-1}$, and
therefore, it is natural to expect some deviations from the spectra found in standard
treatments. In order to illustrate the nature of these deviations  we will consider a simple modification from the above mentioned
pattern which we parameterize here by $\eta_{\nk}^c = A/k + \beta$, where $\beta$ a constant with dimensions of length such that $-\infty < s_{\vec{k}}^c < 0$. When $\beta=0$ we recover a flat spectrum. 
For the sake of exploring the dependance on $k$ in the power spectrum as given by \eqref{power}, it is convenient to define the dimensionless quantity $x \equiv kR_D$ (recall that $R_D \simeq 4000$ Mpc).  If we assume that the time of collapse of each mode is given by $\eta_{\nk}^c = A/k + \beta$, then $s_{\nk}^c (x) \equiv k \eta_{\nk}^c = A + Bx$, with $B \equiv \beta/R_D$. In this way $A$, $B$ and $x$ are dimensionless quantities. It is important to note that the multipoles $l$ in the observed angular power spectrum cover the range $2 \leq l \leq 2600$, which corresponds to the modes with $10^{-3}$ Mpc$^{-1}$ $\leq k \leq$ 1 Mpc$^{-1}$, that is, the range of observational interest for $x$ is $1 \leq x \leq 10^{3}$. Furthermore, for the modes which the Sachs-Wolfe effect is dominant, the corresponding range of $x$ is given by $1 \leq x \leq 20$.

Once we have chosen a particular form for $\eta_{\nk}^c$, we can compute the value of the scale factor at the collapse time $a^c_k \equiv a(\eta_{\nk}^c)$, and compare it with the traditional value of the scale factor at the time of ``horizon crossing" during the inflationary regime, $a^H_k \equiv a(\eta_{\nk}^H)$, where $\eta_{\nk}^H$ is the conformal time of horizon crossing for a mode $\nk$ during inflation. The horizon crossing occurs when the length corresponding to the mode $k$ has the same value as the Hubble radius, $H_I^{-1}$, i.e. when $k=aH_I$ for comoving modes; therefore, $a_k^H =k/H_I$ (we recall that during the inflationary stage $H_I$ can be considered as a constant). Thus, the ratio of the value between the scale factor at horizon crossing for a mode $k$ (during the inflationary regime) and its value at the time of collapse for the same mode is

\begin{equation}\label{as}
\frac{a_k^H}{a_k^c} \simeq \frac{k}{H_I} \bigg( \frac{-1}{H_I \eta_{\nk}^c} \bigg)^{-1} = k |\eta_{\nk}^c| = |A + Bx| = |s_{\nk}^c|.
\end{equation}
Thus, for every mode  $\nk$, we can  read  directly  from  this  equation and   our
parametrization for the collapses (in terms of  $A$, $B$ and $x$), the relationship
between  the  scale factor  at collapse and  at horizon crossing.

\begin{figure}
  \begin{center}
  \includegraphics{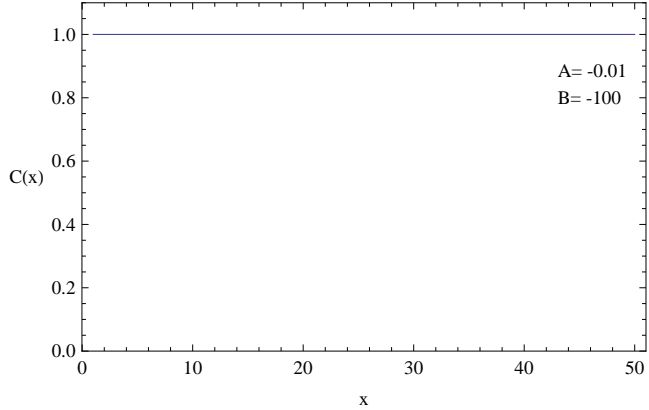}
  \end{center}
\caption{The function $C(x)$ in the interval $x \in [1,50]$, with $A=-0.01, B=-100$ and $\lambda_v = \lambda_{\pi} = 1$. In this case $a_k^H = |s_{\nk}^c| a_k^c$ with $|s_{\nk}^c| \in [100.01, 5000.01]$, i.e. $a_k^c \ll a_k^H$.}
\label{fig:6}       
\end{figure}

It is interesting to note that we can  recover a {\it nearly} flat power spectrum if we demand that, within the symmetric scheme
$\lambda_v=\lambda_\pi =1$, the collapses take place at $s_{\vec{k}}^c\to -\infty$. 
In that case (and up to the first order in $1/s_{\vec{k}}^c$) the expression (\ref{ps}) takes the form
\begin{equation}\label{ps-infty}
\mathcal{P}_{\psi}(k)_{s_{\vec{k}}^c \to -\infty} \equiv \epsilon\frac{t_p^2H^2}{8\pi^2} \left[1+\mathcal{O}^2
\left(\frac{1}{s_{\vec{k}}^c}\right)\right].
\end{equation}
 That is,  in contrast to what is often  assumed in the standard approach to  the problem,  this
  option  would   correspond to the  collapses (the process that breaks the initial symmetry,  usually considered as tied to the  ``the quantum  to classical transition") taking place when the modes are  ``well inside the Hubble radius" see \eqref{as} above. This situation is illustrated in Fig. \ref{fig:6} for some special values of the parameters $A$ and $B$, and assuming the symmetric collapse scheme. This  is one case  where  no deviation from a flat power spectrum is observed.

\begin{figure}
\begin{center}
  \includegraphics{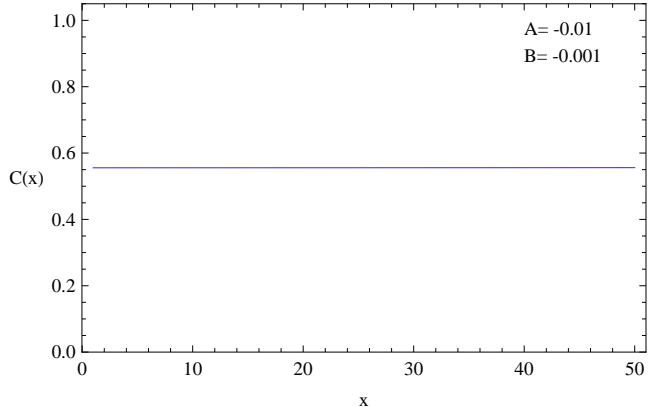}
  \end{center}
\caption{The function $C(x)$ in the interval $x \in [1,50]$, with $A=-0.01, B=-0.001$ and $\lambda_v = \lambda_{\pi} = 1$. In this case $a_k^H = |s_{\nk}^c| a_k^c$ with $|s_{\nk}^c| \in [0.011, 0.06]$, i.e. $a_k^c \gg a_k^H$.}
\label{fig:3}       
\end{figure}

It is also convenient to analyze the limit $s_{\vec{k}}^c \to 0$, corresponding to the opposite case in which the collapses take
place well outside the Hubble radius (this could correspond, for instance, to the reheating time).\footnote{We thank Prof.  R. M.
Wald for pointing this out.} In this case the expression \eqref{ps} takes the form
\begin{equation}\label{pslim}
\mathcal{P}_{\psi}(k)_{s_{\vec{k}}^c \to 0} = \epsilon\frac{t_p^2H^2}{8\pi^2} \left[ \lambda_{\pi}^2 \left(\frac{1}{9} + \mathcal{O}^2
(s_{\vec{k}}^{c}) \right) + \lambda_{v}^2 \left(\frac{2}{3} + \mathcal{O}^2 (s_{\vec{k}}^{c}) \right) \right].
\end{equation}
This case is illustrated in Fig. \ref{fig:3}, and corresponds  to another case  where no deviation from a flat power spectrum is observed.
Thus, even though $s_{\vec{k}}^c$ may have a non-trivial $\vec{k}$ dependence, the spectrum becomes independent of $\vec{k}$, up to small corrections of order $s_{\vec{k}}^c \ll 1$.

\begin{figure}
\begin{center}
  \includegraphics{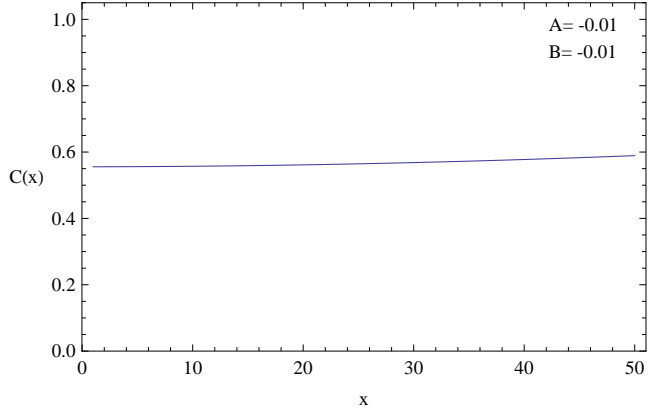}
  \end{center}
\caption{The function $C(x)$ in the interval $x \in [1,50]$, with $A=-0.01, B=-0.01$ and $\lambda_v = \lambda_{\pi} = 1$. In this case $a_k^H = |s_{\nk}^c| a_k^c$ with $|s_{\nk}^c| \in [0.02, 0.51]$, i.e. $a_k^c > a_k^H$.}
\label{fig:2}       
\end{figure}

\begin{figure}
\begin{center}
  \includegraphics{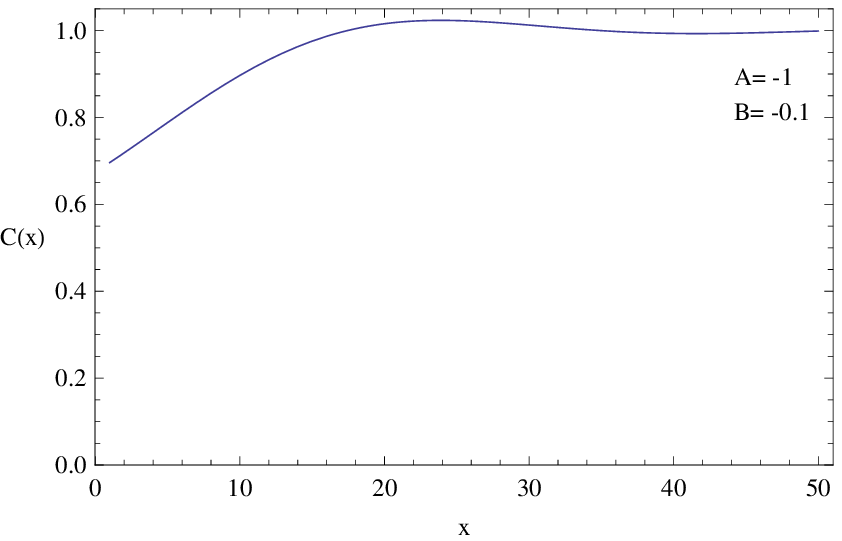}
  \end{center}
\caption{The function $C(x)$ in the interval $x \in [1,50]$, with $A=-1, B=-0.1$ and $\lambda_v = \lambda_{\pi} = 1$. In this case $a_k^H = |s_{\nk}^c| a_k^c$ with $|s_{\nk}^c| \in [1.1, 6]$, i.e. $a_k^c < a_k^H$.}
\label{fig:4}       
\end{figure}

\begin{figure}
\begin{center}
  \includegraphics{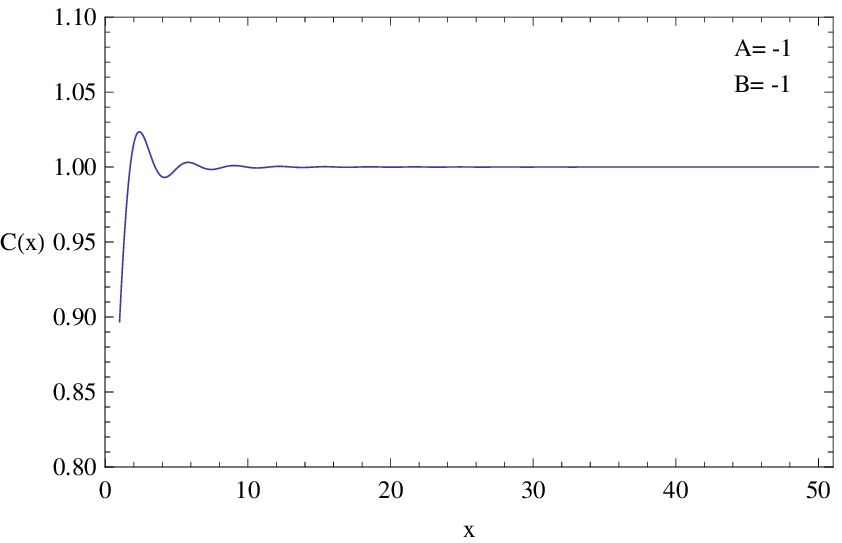}
  \end{center}
\caption{The function $C(x)$ in the interval $x \in [1,50]$, and $A=-1, B=-1$ and $\lambda_v = \lambda_{\pi} = 1$. In this case $a_k^H = |s_{\nk}^c| a_k^c$ with $|s_{\nk}^c| \in [2, 51]$, i.e. $a_k^c < a_k^H$.}
\label{fig:5}       
\end{figure}

We also present the behavior of the function $C(x)$ within the symmetric scheme in some intermediate cases: Fig. \ref{fig:2}, Fig. \ref{fig:4} and Fig. \ref{fig:5}. It is interesting to point out that in Fig. \ref{fig:2}, the function $C(x)$ would induce a pattern in the power spectrum similar to that usually described in terms of a \emph{spectral index} $n_s$, i.e. so that the power spectrum is proportional to $k^{n_s-1}$, with $n_s \neq 1$. Meanwhile, in Fig. \ref{fig:4} and Fig \ref{fig:5}, we observe that the collapse of the wave-function would affect only to the large scale modes $x = kR_D \leq 20$ in both cases (a similar result, in a different context, was found in \cite{Leon:2011hs}). The door is clearly open for a  more thorough investigation of the observationally allowed ranges, which requires direct comparison with data. In order to do that one must include the effects of
the plasma acoustic oscillations, and a detailed analysis similar to that carried out in \cite{Landau:2011aa}, which is beyond the scope of the present work.

We  should point out that  the results  we  present  here should be  relevant  irrespective of the  views  one takes on  our  collapse proposals.
The point is  that, even  if  one follows  the  standard   views  on the matter, but  does  so  in  a self-consistent way,  one  would  come at the  end  to  essentially   the    predictions  above  regarding the form of  the spectrum.   Let us assume that one chooses to ignore the shortcomings of the standard accounts and accepts that,  say decoherence, addresses (somehow)
the issue at hand (i.e.  evades  the conceptual problems discussed in Section \ref{intro} and the references \cite{Perez:2005gh,Sudarsky:2009za}),
   and that the mystery  lies only  in the question  concerning the   precise  mechanism that lies behind the fact that, from the
  ``options"   one   finds  in the  decoherence  analyses (i.e.  those displayed in the  reduced   density matrix), a  single  particular one   seems
   to be  selected  for ``our  branch".  Within this  point of  view,  one  would  be  assuming that   the initial  symmetry  has  been lost $-$at least
for practical  purposes (i.e. in our branch)  as the  relevant   situation  would not longer  be  described  by the full  fledged superposition of  inhomogeneous  and  isotropic  states (that  make  up the  DB vacuum)  but  by the  state  corresponding to our branch,  as presumably one  would  be advocating  when adopting  such position$-$. That is,  we    would need to focus on a particular state that  corresponds  to  the  particular
realization or actualization (represented  by  a particular  element in the  density matrix). Thus, it seems  clear that for the sake of
 self-consistency, when  studying  aspects of the anisotropies in the  CMB,  one should  consider that state  corresponding to
such ``selected  option",  and  not the entire   vacuum state  which  describes the homogeneous and isotropic state of affairs previous
to the ``selection".\footnote{The  selection of course  refers to the fact that,  according to the  standard  arguments,  the
resulting density matrix,  after  becoming  essentially diagonal due to  decoherence, represents  an  ensemble  of universes, and
our  particular one  corresponds to one  of them.  That  one  can be
  considered  as  selected  by Nature  to become realized. Alternatively, one might  take the view  that  these other universes
are also  realized, and thus  they  also exits  in  realms completely inaccessible to us. In that case  the selection  corresponds to
that universe  in which  we happen to  exist.} In following such views,
the  discussion  that we are presenting in this paper
 would  have  to be taken to  represent  the effective description corresponding
to ``our  perceived  universe"  (in  a context  where
one   puts together something like the   many-worlds interpretation  with  the
arguments based on   decoherence).
Although we definitely do not adhere such view for the reasons explained in
\cite{Sudarsky:2009za}, it is clear that an effective description such as
the one presented here is what would have to be contemplated when
dealing with the issues within any view which pretends to allow one to deal with
the details characterizing the inhomogeneities and anisotropies in the cosmic
structure and its imprints in the CMB that we do observe.


Finally, it is worth mentioning that in the collapse model introduced in \cite{Perez:2005gh}, where the collapse was considered within
the framework of semiclassical gravity, one could also obtain a similar expression for the power spectrum to that given in equation (\ref{power}).
In that case only the matter fields were quantized, with the space-time metric considered to be an effective (classical) description
of gravity. Using semiclassical gravity one obtains
\begin{equation}\label{pscol}
\mathcal{P}_{\psi}(k) = \epsilon\frac{t_p^2H^2}{8\pi^2} \left[ \lambda_{1}^2 \left(1+\frac{1}{(s_{\vec{k}}^c)^2}\right) \sin^2 s_{\vec{k}}^c + \lambda_2^2 \left( \cos s_{\vec{k}}^c - \frac{\sin s_{\vec{k}}^c}{s_{\vec{k}}^c}  \right)^2 \right],
\end{equation}
with $\lambda_1$ and $\lambda_2$ two real numbers analogous to those given in $\lambda_\pi$ and $\lambda_v$.
Again, assuming $\lambda_1=\lambda_2=1$ and taking the limit when $s_{\vec{k}}^c \to -\infty$ we arrive to an expression of the form (\ref{ps-infty}). However, taking the
limit $s_{\vec{k}}^c \to 0$ in \eqref{pscol} we obtain
\begin{equation}\label{pscol2}
\mathcal{P}_{\psi}(k)_{s_{\vec{k}}^c \to 0} = \epsilon\frac{t_p^2H^2}{8\pi^2} \left\{ \lambda_{1}^2 \left[ 1+\mathcal{O}^2 (s_{\vec{k}}^{c})
\right] + \lambda_2^2 \left[ \mathcal{O}^4 (s_{\vec{k}}^{c}) \right] \right\}.
\end{equation}
Note that expression \eqref{pscol2}
exhibits a distinct behaviour from \eqref{pslim}. That is, if the collapse scheme is such that $\lambda_1=0$ and $\lambda_2=1$
(referred to as the Newtonian scheme in \cite{Perez:2005gh}), then the shape of the spectrum is not flat
but proportional to terms of order $(s_{\vec{k}}^c)^4$. This contrasts with the behaviour of  \eqref{pslim},
where regardless  of  the values of $\lambda_v$ and $\lambda_\pi$, the spectrum is always flat plus small corrections of order $(s_{\vec{k}}^c)^2$.

In  other  words,  the predictions regarding  the  shape of the spectrum depend
strongly on  what   is the variable that characterizes the collapse,
and the times  at which the collapse  of each mode takes place. The effect
becomes  substantially reduced if  the   collapse  is tied to  the  Mukhanov-Sasaki
  variable, in the  approach investigated in this work,  or if it is tied to the field  variable,
in the approach  studied in previous works, while the effects would be generically very large in the case where the collapse is tied to the momentum conjugate field variable in that approach. On the other hand, the spectrum would become close to the standard one if the collapse takes place  always   with  very small values of $s_{\vec{k}}^c$.

The point is that,  even if  the  appropriate   treatment of  the
situation at hand  involves  either a  collapse in the Mukhanov-Sasaki  variables,  or  a  collapse  in the  inflaton field,
 we face the prospect of important deviations (from the  flat one) in the
predictions of the primordial power spectrum. The  fact  is that in any scheme thought
to be controlled by  essentially stochastic  rules,  one  would not expect  any
relationship $-$like that requiring the  collapse  to  occur   always for   very
small values of $s_{\vec{k}}^c$, or  in a way that  this  quantity   was  always
independent of $\vec{k}$$-$ to hold  exactly  for all  the  modes involved,
and thus interesting departures from the  standard  predictions  are to be expected. Needless is to say that  the  effects  might
 be much more important in the event  that the  collapse  was  more appropriately treated   with  the  Newtonian  scheme  (in which the collapse  occurs   in the momentum  conjugate to  inflaton field  modes).
The discussion above shows how the parameters characterizing the different collapse schemes lead to predictions which can, in principle,  be compared with the observational data, and are different from the standard ones even when we followed the standard approach in quantizing the gravitational sector of the cosmological perturbations.

\section{Discussion}\label{discussion}

In this paper we have analyzed to what extend do  the general properties and details of the results  obtained in previous  works
 (e.g. see \cite{Perez:2005gh,DeUnanue:2008fw})
depend on the specific approach one takes  to deal with the quantum  aspects of the problem.
In those  original treatments  the analysis   of  the collapse  had  relied on what is commonly known  as  semiclassical  gravity:
a classical  description for the space-time metric  (including its  perturbations) coupled  to  a quantum treatment of the
 inflaton field. 
The collapse was  assumed  to  affect   the state of the inflaton field, while  the metric   simply   ``back-reacts" to the change in the  expectation value
of the energy-momentum tensor, leading to  a geometry that is no longer homogeneous and isotropic. %
In the  present  work we implemented the collapse hypothesis  on a variable  that simultaneously  characterizes the metric and the
scalar field  perturbations at the quantum level, the so  called Mukhanov-Sasaki variable. That is,  we  simply  incorporated
the collapse  hypothesis  into the standard treatment found  on the  literature on the subject,  as a way  to deal with the  basic
issue of the transition from the homogeneous and isotropic situation to another one lacking those symmetries,  and thus  containing the seeds  of  cosmic  structure.
We have  seen  that  confronting this  issue  leads one to  results  indicating that one can  generically  expect   deviations  from the flat   primordial spectrum.
We have also argued that    even if  one  decides  to  ignore its  shortcomings  and
adopt the  standard  posture   in  which  decoherence  plus the many  world
interpretation  is   taken to  address the  emergence of  inhomogeneity and
anisotropy,  simple  self-consistency  would lead  one to  find  essentially the
same  deviations  in the  form  of the power spectrum  as  we have found here.

Going back to the present work, the point of view taken here a contrast
with that of our previous studies in the specific variable one takes for the
realization of the collapse hypothesis.
  In both types of  treatments (the  previous ones relying on semiclassical gravity and the present one relying on the joint quantization
of metric and  inflaton perturbations) we have found that the details of the specific model for the collapse
do have an  impact on the
CMB  power spectrum.  We  have found that   similar results   are  obtained
in both approaches if one   assumes that the  collapse  occurs  always for very
small values of $s^c_{\vec{k}}$, and one  avoids the purely Newtonian  scheme of
our  previous works. On the other hand,  it is  easy to see that one of the most
important differences   between the two approaches   refers to  the
 predictions  on the existence (or lack)  of primordial  tensor modes  generated  by  the  exact  same  mechanism
(and thus  at comparable  magnitude  level) as the scalar ones.
This  leaves an important open question: which  one  of  the two  is
the most appropriate  treatment of the subject?

We conclude by briefly discussing  our current views on this  issue (the interested reader is  directed to Section 8 of reference \cite{Sudarsky:2009za}).
 Let  us consider how can the ``collapse of the wave function" fit into our  general understanding of physical theory.  The basic  constituents of  our world,  as far as we understand them now,  are the matter fields  described  by the standard  model of particle  physics  (augmented to incorporate the  masses of  neutrinos and  the inflaton),  the  still mysterious  dark matter and dark energy, and the gravitational sector. Of these,  the  standard fields (including the electroweak and strong interactions, as well as the  quarks and leptons, and the inflaton)  seem to fit quite nicely into a framework incorporating   quantum  field theory on any reasonable  background space-time. The  dark matter is likely to be   described by some  other fields with a similar structure as that given for the ordinary  matter (although  perhaps involving  distinct novel aspects  like supersymmetry), and dark energy seems to be most economically   described by a  cosmological constant (although
   certainly naturalness and  other kinds  of  issues  are still outstanding). However,  the component of  our   world which
seems hardest to fit  with the   general  paradigms offered by quantum theory is  gravitation.

There  exist  a  very extensive literature on this subject and  we  will not even  attempt to describe all the  problems,   either technical or conceptual, found in this road. However,   it seems  quite clear  that conceptually there  is  room for  large  differences  from the usual cases  to arise  when considering  the incorporation of the quantum aspects of Nature  in the gravitational  context. 
According to general relativity, gravitation reflects the structure of space-time itself, whereas  quantum theory seems to  fit most easily in  contexts  where this  structure is  a given one.  That  is,
quantum states   are associated  with
 objects that ``live" in space-times. For instance,   the standard  Schr\"odinger equation   specifies the time  evolution  of a system, the quantum states of fields  characterize   the system in connection to algebras of
 observables associated with predetermined  space-time regions, and so forth.  It  is clear that mayor  conceptual modifications are in order  if we want  to  describe  the space-time itself  in  a quantum language.
  This issue appears  in various  guises  in  the  different approaches now available to quantum  gravity, most conspicuously  as  the problem of time  which afflicts all   attempts to  deal with the subject following a canonical approach.
 It seems thus  natural to   speculate that it  is  precisely in this  setting where something that departs as dramatically from the  quantum orthodoxy as the dynamical collapse  of the wave function might find  its   origin.
 That is,  it seems plausible that in  dealing with  such  conceptual   problems,
 fundamental obstacles  might prevent the   emergence of   the   usual   quantum theory as the full effective description when gravity is concerned.
 Lingering aspects of that more fundamental description would  take the form of deviations from the standard unitary  evolution that characterize quantum theory as we know it.  In fact, there are  already indications
  about such  deviations  in  analysis that attempt to recover  time, in a  relational  setting,  by   using some   variables of the
theory  to play the role of physical clocks (see for instance the references \cite{Pullin:2004,Pullin:2004a}).

In other   words, it seems  natural  to  conjecture that  the departure form the standard paradigm,
that we  have considered  here  as described  effectively by the ``collapse of the wave function", corresponds to lingering features  of  the  fundamental timeless (and  probably spaceless) theory  of quantum gravity. If that is the case, the emergence of space-time itself would  be tied to the incorporation of  such  effective quantum description of matter  fields living on  space-time, and  evolving  approximately according to  standard quantum field theory on  curved spaces, with some small  deviations which might  include our hypothetical collapse.  In   that context   it seems  clear that space-time itself would  be  nothing but an effective description
  of the underlying quantum gravity reality. Ideas of  this sort regarding   emergent gravity have been  indeed considered  previously,  for instance in \cite{Jacobson:1993,Ashtekar:2001}. This  suggests that in the context where we consider the collapse of the wave function the space-time itself must be regarded as an approximate phenomenological description, and thus  as something that can {\it not} be subjected to quantization. Let us imagine for a moment that we  want to consider  the propagation of heat in a medium. It  is  well known that this can be  described by the heat equation $\partial T/ \partial t - \nabla^2 T = S $,  where $S$ represents the  heat sources. It is quite clear that despite the fact that this looks
  like a standard type of equation for some field, it would be meaningless to quantize it.
    Moreover, we can imagine  some situation  in  which the source of heat  requires  a quantum mechanical  treatment, so that $S$  becomes some quantum operator.
    Under such conditions   it seems reasonable that  to the extent that   the   temperature description is  still relevant and of interest,  the right hand side of the equation above
    should be replaced by something like $\langle \hat S \rangle$.
Of course there will be situations  that are so far removed  from the  context
where the heat  equation was derived   that  even the notion of temperature itself
  would  become meaningless.  We  equally expect that in the quantum gravity
theory we will be able to find  many    situations where the semiclassical
Einstein's equations would   be  completely inappropriate,  but  in following with
our line of thought and simple analogy above, it seems quite likely that those would correspond to situations where the concept of   space-time itself  becomes   meaningless.

Of course all these arguments above are  filled with ``educated" guesses and
conjectures, and we can not take them  as more than  a guidance. Then, it is
important to  determine  to what  extent our predictions depend on the precise
way to  implement our ideas: semiclassical gravity or the Mukhanov-Sasaki approach.
The  study  of   this  question  was  the main purpose of the analysis  we have
carried out in this manuscript.
We have found that the  predictions made by our proposal (i.e. the collapse) are
generically different from the standard ones, and can be directly confronted with observations. (Indeed, as it has been argued in this paper, and was previously noticed in \cite{Perez:2005gh,Sudarsky:2009za}, the standard treatment do not make any prediction at all).
 On the other hand, the predictions found in this paper (and  which   were obtained following the standard approach of quantizing the perturbations in the geometry but with the additional ingredient we called the self-induced collapse hypothesis) 
 are very similar to those obtained in previous works (using the semiclassical approach) as far as the scalar perturbations are concerned. This strongly suggests that it is not the manner in which we treat the gravitational interaction (truly fundamental or as in an effective theory) that leads to predictions different to the standard ones, but the collapse hypothesis itself, which was introduced  into the treatment in order to  deal with the more fundamental issues we touched at the Introduction regarding the origins of the primordial perturbations.

\section*{Acknowledgments}

 We are  glad  to acknowledge  very useful  discussions with Prof. R. M. Wald.
 The work of ADT is supported by a UNAM postdoctoral fellowship and the   CONACYT  grant No 101712. The work of GL and DS is  supported  in part by the   CONACYT grant
No 101712. GL acknowledges financial support by CONACYT postdoctoral grant. DS  was supported also  by   CONACYT  and  DGAPA-UNAM   sabbatical  fellowships,  UNAM-PAPIIT IN107412-3 grant, and gladly acknowledges the IAFE-UBA  for  the  hospitality  during a sabbatical stay.

 \appendix

 \section{$A_{\vec{k}}(\eta,\eta_{\vec{k}}^c)$, $B_{\vec{k}}(\eta,\eta_{\vec{k}}^c)$, $C_{\vec{k}}(\eta,\eta_{\vec{k}}^c)$ and $D_{\vec{k}}(\eta,\eta_{\vec{k}}^c)$ in equation (\ref{v.y.pi})}\label{A,B,D,D}

In a universe close to de Sitter the functions $A_{\vec{k}}(\eta,\eta_{\vec{k}}^c)$, $B_{\vec{k}}(\eta,\eta_{\vec{k}}^c)$, $C_{\vec{k}}(\eta,\eta_{\vec{k}}^c)$ and $D_{\vec{k}}(\eta,\eta_{\vec{k}}^c)$ in equation (\ref{v.y.pi}) take the form
\begin{eqnarray}
X_{\vec{k}} (\eta,\eta_{\vec{k}}^c)&=& X_{\vec{k}}^{(1)} (\eta,\eta_{\vec{k}}^c)\cos \Delta_{\vec{k}}^c + X_{\vec{k}}^{(2)} (\eta,\eta_{\vec{k}}^c)\sin \Delta_{\vec{k}}^c.
\end{eqnarray}
Here $X$ denotes $A$, $B$, $C$ or $D$, with the different $X_{\vec{k}}^{(i)}(\eta,\eta_{\vec{k}}^c)$ given by
\begin{subequations}
\begin{eqnarray}
A_{\vec{k}}^{(1)} = 1-\frac{1}{s}\,C_{\vec{k}}^{(1)} ,\quad \hspace{.15cm}&&
A_{\vec{k}}^{(2)} = \frac{1}{s_{\vec{k}}^c}-\frac{1}{s}\,C_{\vec{k}}^{(2)} ,\\
B_{\vec{k}}^{(1)} =-C_{\vec{k}}^{(1)}C_{\vec{k}}^{(2)} ,\quad \hspace{.18cm}&&
B_{\vec{k}}^{(2)} =- 1+\frac{1}{s^2}+\frac{1}{s_{\vec{k}}^{c\,2}}-\frac{1}{ss_{\vec{k}}^c}C_{\vec{k}}^{(2)},\\
C_{\vec{k}}^{(1)} = \frac{1}{s}-\frac{1}{s_{\vec{k}}^c},\hspace{.7cm}\quad &&
C_{\vec{k}}^{(2)} = 1+\frac{1}{ss_{\vec{k}}^c},\\
D_{\vec{k}}^{(1)} = 1+\frac{1}{s_{\vec{k}}^c}\,C_k^{(1)},\quad &&
D_{\vec{k}}^{(2)} =-\frac{1}{s}\left(1-\frac{s}{s_{\vec{k}}^c}\,C_{\vec{k}}^{(2)}\right).
\end{eqnarray}
\end{subequations}
Remember that we are using $s\equiv k\eta$, $s_{\vec{k}}^c\equiv k\eta_{\vec{k}}^c$ and $\Delta_{\vec{k}}^c\equiv s-s_{\vec{k}}^c$.



\end{document}